\documentclass[%
reprint,
nofootinbib,
 amsmath,amssymb,
 aps,
superscriptaddress,
showkeys,
longbibliography
]{revtex4-1}

\usepackage{xr}
\usepackage{tabularx}
\usepackage{graphicx}
\usepackage{dcolumn}
\usepackage{bm}

\makeatletter
\newcommand*{\addFileDependency}[1]{
  \typeout{(#1)}
  \@addtofilelist{#1}
  \IfFileExists{#1}{}{\typeout{No file #1.}}
}
\makeatother

\newcommand*{\myexternaldocument}[1]{%
    \externaldocument{#1}%
    \addFileDependency{#1.tex}%
    \addFileDependency{#1.aux}%
}

\myexternaldocument{supplementary_material}

\begin{document}

\preprint{APS/123-QED}

\title{An Analysis of the Influence of Transfer Learning When Measuring the Tortuosity of Blood Vessels}

\author{Matheus V. da Silva}
\affiliation{Department of Computer Science, Federal University of S\~ao Carlos, S\~ao Carlos, SP, Brazil}

\author{Julie Ouellette}
\affiliation{Department of Cellular and Molecular Medicine, Faculty of Medicine, University of Ottawa, Ottawa, ON, Canada\looseness=-1}
\affiliation{Neuroscience Program, Ottawa Hospital Research Institute, Ottawa, ON, Canada}

\author{Baptiste Lacoste}
\affiliation{Department of Cellular and Molecular Medicine, Faculty of Medicine, University of Ottawa, Ottawa, ON, Canada\looseness=-1}

\author{Cesar H. Comin}
\email[Corresponding author: ]{chcomin@gmail.com}
\affiliation{Department of Computer Science, Federal University of S\~ao Carlos, S\~ao Carlos, SP, Brazil}

\date{November, 2021}

\begin{abstract}
Convolutional Neural Networks (CNNs) can provide excellent results regarding the segmentation of blood vessels. One important aspect of CNNs is that they can be trained on large amounts of data and then be made available, for instance, in image processing software. The pre-trained CNNs can then be easily applied in downstream blood vessel characterization tasks, such as the calculation of the length, tortuosity, or
caliber of the blood vessels. Yet, it is still unclear if pre-trained CNNs can provide robust, unbiased, results in downstream tasks when applied to datasets that they were not trained on. Here, we focus on measuring the tortuosity of blood vessels and investigate to which extent CNNs may provide biased tortuosity values even after fine-tuning the network to the new dataset under study. We show that the tortuosity values obtained by a CNN trained from scratch on a dataset may not agree with those obtained by a fine-tuned network that was pre-trained on a dataset having different tortuosity statistics. In addition, we show that the improvement in segmentation performance when fine-tuning the network does not necessarily lead to a respective improvement on the estimation of the tortuosity. To mitigate the aforementioned issues, we propose the application of data augmentation techniques even in situations where they do not improve segmentation performance. For instance, we found that the application of elastic transformations was enough to prevent an underestimation of 8\% of blood vessel tortuosity when applying CNNs to different datasets. 
\end{abstract}


\maketitle

\section{Introduction}

Blood vessels take part in many physiological processes in humans and animals and can be found almost anywhere in an organism. Therefore, many ailments and developmental disorders may be caused by abnormal blood vessels \cite{Potente2011}. Thus, characterizing blood vessels is an important matter not only for diagnosis but also to help answer important research questions regarding angiogenesis \cite{Guzel2020, Fernandez-Klett2020}, blood vessel related ailments \cite{Canton2021} and the blood-brain barrier \cite{Fernandez-Klett2020, FriasAnaya2021}. Common metrics for characterizing blood vessels are density and tortuosity since they have been shown to influence neuronal activation \cite{Lacoste2014} and blood flow \cite{Han2012}. Acquiring precise values for those metrics usually requires annotating blood vessels in images, which is time-consuming and error-prone. Consequently, many image processing techniques have been developed for automatically characterizing blood vessels \cite{Ma2019, Tongpob2019}. Usually, the most challenging step of the developed techniques lies in segmenting the blood vessels \cite{Lesage2009}.

More recently, owing to the recent advancements of Deep Learning methods on many computer vision tasks \cite{Goodfellow2016}, Convolutional Neural Networks (CNN) have been applied with great success for segmenting blood vessels. Among the most common types of images where CNNs have been applied for blood vessel analysis are retina fundus and magnetic resonance images \cite{Li2021, Moccia2018}. Recent works have also explored the use of CNNs for blood vessel microscopy images \cite{Tetteh2020, Kirst2020, Todorov2020}. Many CNN architectures have been defined for blood vessel segmentation, U-Net-based architectures \cite{Li2021} being the most common. Given that annotating blood vessel images, which can contain thousands of blood vessel segments, is a demanding task, most approaches for training CNNs involve transfer learning methods \cite{Tetteh2020, Kirst2020, Todorov2020}. This is especially true for 3D images, since tracing blood vessels in volumetric images is a particularly difficult task. Most commonly, transfer learning consists in obtaining some off-the-shelf network, that may or may not have been trained on blood vessel images, and fine-tuning the network for segmenting blood vessels for the type of image under study.

\begin{figure}
    \includegraphics[width=\columnwidth]{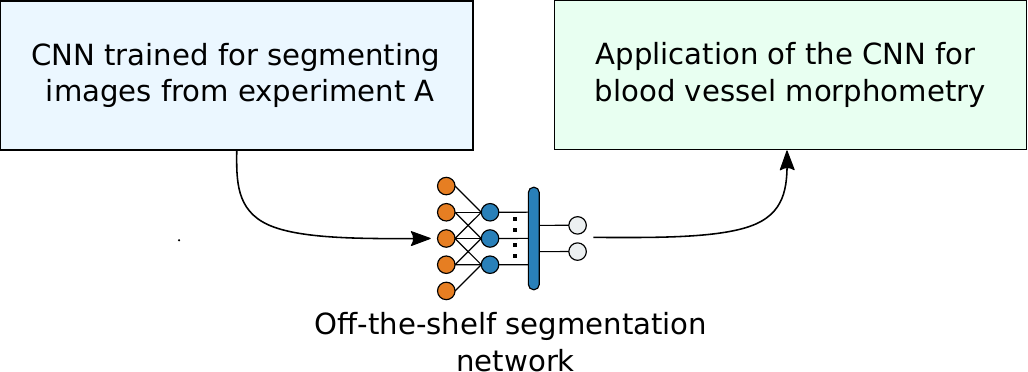}
  \caption{Illustration of a transfer learning approach commonly used for training CNNs. A CNN is trained for image segmentation and made available to other researchers. The off-the-shelf CNN is then used as is, or fine-tuned, on another dataset and included in a pipeline for blood vessel morphometry.}
  \label{f:fig_intro}
\end{figure}

In this work, we investigate a possible pitfall when using transfer learning for blood vessel segmentation. The following situation, illustrated in Figure~\ref{f:fig_intro}, is considered. Suppose that a CNN was trained based on a set of images that were obtained and manually annotated from some clinical trial or biological experiment. Most likely, the objective of the training was to maximize some segmentation performance metric such as the Intersection over Union (IoU) or the Dice coefficient. The trained network was then made available on some repository or software. Then, a new set of images obtained under distinct experimental conditions was acquired during some other experiment. Notice that the experiment might even be carried out by a different group of researchers. Suppose that the overall objective of this new experiment is to study the tortuosity of the blood vessels. For this task, the off-the-shelf network trained using the data from the previous experiment is applied for segmenting the blood vessels, followed by a calculation of the tortuosity. 

The above situation has two important possible drawbacks that need to be taken into account. First, the objective of the original experiment where the network was initially trained was solely the segmentation of the blood vessels, there is no guarantee that the values obtained for the tortuosity are correct. Second, the data used in the two situations might have distinct characteristics. Here we do not focus on possible changes on the image characteristics, such as brightness or contrast, or the type of blood vessel under study, but on more subtle differences that might affect the blood vessels. Specifically, we consider two sets of images that have similar statistics but contain blood vessels having distinct degrees of tortuosity. The idea being that a CNN trained on blood vessels having low tortuosity, when used for segmenting highly tortuous blood vessels, might lead to an overestimation or underestimation of the tortuosity values. The same situation is studied for the opposite problem, that is, CNNs trained on tortuous blood vessels being used for segmenting non-tortuous blood vessels. 

This study is particularly relevant for current researches involving blood vessels, where the impact of an experimental condition might lead to subtle changes in the morphology of the blood vessels, which need to be quantified using robust and systematic approaches. We specifically focus on the tortuosity in this study because it is a difficult metric to be measured and compared by visual inspection, and thus requires particularly reliable means for automatic quantification. 

Two main investigations are considered. The first is to assess if there are indeed possible biases that may appear on the calculated tortuosity when using off-the-shelf CNNs. Provided this is true, we verify to which extent such biases can be avoided by fine-tuning the network on the new dataset. This requires the manual annotation of additional images, and thus we also verify how many images need to be annotated in order to reduce the observed biases. In addition, we also investigate a possible data augmentation technique that can be used when training the networks in order to better generalize them for segmenting blood vessels possessing different degrees of tortuosity. All in all, our analysis involves training a CNN over 7000 times in different experimental conditions.

The analysis is focused on a large dataset of confocal microscopy images, but the results should be general for other types of images having blood vessels with a similar appearance.

\section{Literature Review}

As stated by \textcite{Moccia2018}, traditional approaches for blood vessel segmentation can be divided into three categories: deformable models, tracking, and vessel enhancement. Deformable models comprise segmentation techniques that rely on parametric curves (or surfaces, in 3D) that evolve towards the vessel boundaries~\cite{Al-Diri2009, Zhao2015, Valencia2007, Law2010}. Tracking methodologies are usually based on region-growing techniques for iteratively finding blood vessel pixels~\cite{Mendonca2006, Martinez-Perez2007}. In this case, the set of seed points can be defined either manually or automatically. Blood vessel enhancement approaches aim at reducing image noise and increasing the contrast between the blood vessels and the background. Many enhancement techniques are based on the eigenvalues of the Hessian matrix~\cite{Lorenz1997, Frangi1998, Sato1998}.

Recently, many Deep Learning methodologies have been developed for blood vessel segmentation. Much of the effort has been focused on the segmentation of 2D retinographies. This is likely due to the availability of many annotated public datasets. For instance, ~\textcite{Liskowski2016} evaluated six methodologies for retinal vessel segmentation and found that correcting class imbalances tends to be more critical for performance than applying data augmentation techniques. In the last few years, several retinal vessel segmentation analyses involving different adaptations of the U-Net architecture have been published~\cite{Xiao2018, Zhang2018, Jin2019, Gu2019}. 

One disadvantage of public retinography datasets is that they are usually small. For instance, the DRIVE dataset~\cite{staal2004ridge}, one of the most popular retinography datasets in the literature, contains only 40 images. Thus, several works in the literature use transfer learning for improving the convergence of the networks, which is done by either pre-training with extensive datasets such as the ImageNet \footnote{https://image-net.org/} \cite{Jiang2018, MartinezMurcia2021, Maninis2016}, or by generating additional synthetic data \cite{Zhao2018, Zhao2019, Andreini2019}.


Besides retinography analysis, there have been some important recent developments regarding the segmentation of blood vessels in the brain, such as the DeepVesselNet pipeline \cite{Tetteh2020}. This pipeline addresses the issue of collecting and annotating structures in large volumetric data. The authors suggested applying a pre-training step using a database composed of artificially generated blood vessels. This transfer learning approach allowed a faster network convergence while maintaining a high-quality segmentation. \textcite{Todorov2020} presented a methodology for analyzing important characteristics of whole mouse brain vasculature. The network architecture used was similar to DeepVesselNet, except for the input layer, which used two-channel images. A transfer learning approach involving artificially generated data was also used. With only 0.02\% of the collected mouse brain being annotated, the authors were able to segment the vasculature with high accuracy. \textcite{Kirst2020} also defined a methodology for the segmentation of whole mouse brain vasculature using a combination of traditional methods and CNNs. While capillaries could be segmented by image filtering, thicker vessels appeared as hollow tubes. Thus, a CNN trained with synthetic data was used to fill hollow regions.

The majority of the developed methodologies were aimed mainly at improving the segmentation accuracy of blood vessels. Despite the development of additional performance metrics such as clDice~\cite{shit2021cldice}, the influence of segmentation on downstream tasks, such as blood vessel morphometry, has rarely been addressed. 




\section{Training and fine-tuning off-the-shelf segmentation CNNs}

\begin{figure*}
    \includegraphics[width=\textwidth]{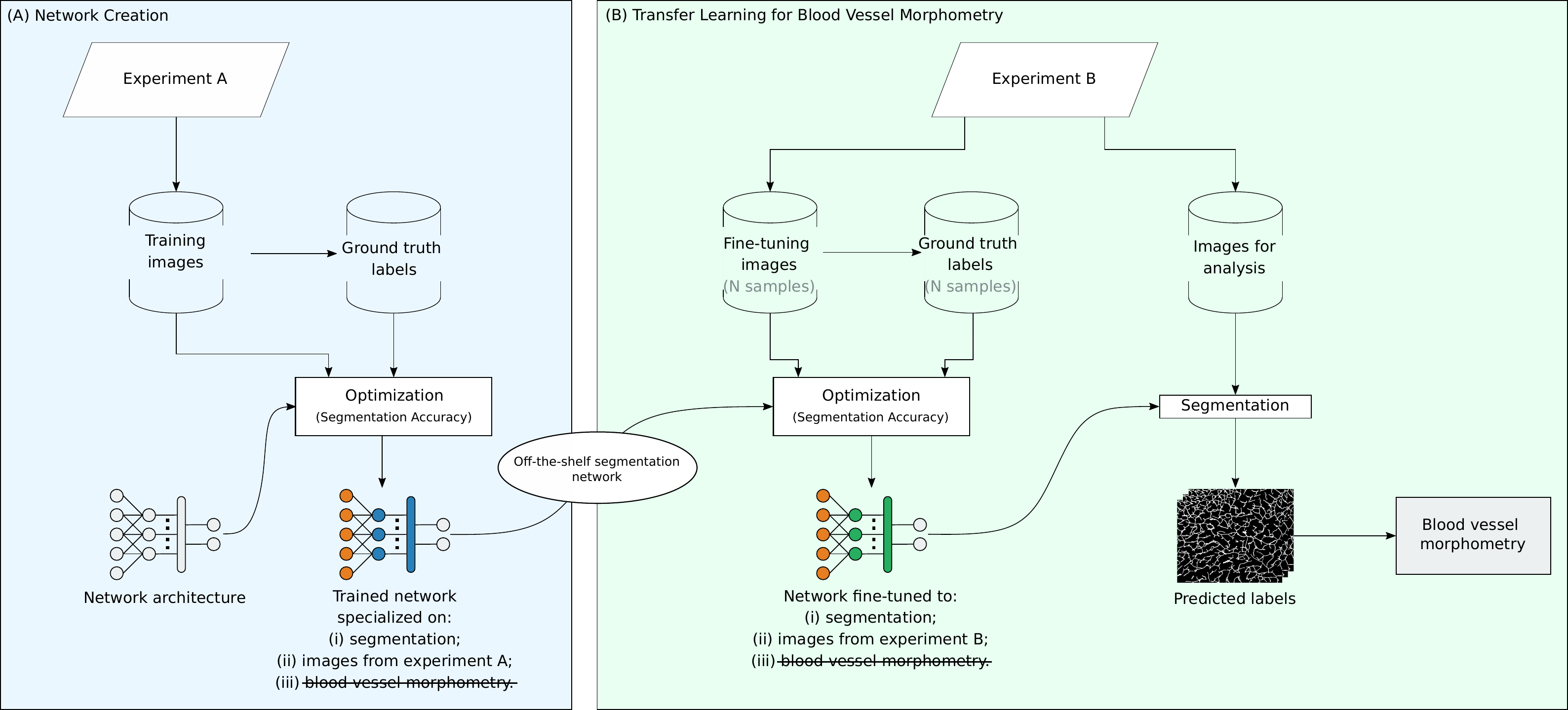}
  \caption{Detailed illustration of the transfer learning approach depicted in Figure~\ref{f:fig_intro}. (A) Images obtained from a specific experimental condition are annotated and used for training a neural network. This network is then made publicly available. The training focus on a segmentation task. (B) The network is used for processing images obtained from another experiment. Additional images are annotated and used for fine-tuning the network using the new data. Morphological properties of the blood vessels identified in the new data are then obtained.}
  \label{fig:motivation}
\end{figure*}

As stated above, transfer learning is an important technique for training CNNs. This is particularly true when the size of the dataset is small or it is costly to annotate the images. Figure~\ref{fig:motivation} illustrates in more detail the common approach for transfer learning depicted in Figure~\ref{f:fig_intro}. First, a group of researchers interested in segmenting blood vessels obtains a set of training images and manually annotates them to optimize a neural network (Figure \ref{fig:motivation}A). After choosing a proper network architecture and optimizing it, a trained neural network is obtained. This neural network is usually evaluated through pixelwise accuracy metrics, such as IoU and Dice. Next, the researchers might make the trained network available on a public source-code repository or an image processing software. 

The pre-trained network can then be used by other research groups in future studies aimed at measuring the morphology of blood vessels. The network might be used as is, or an additional fine-tuning step can be applied to adjust the network to the new data (Figure \ref{fig:motivation}B). For fine-tuning the network, a subset of the new data must be manually annotated. Here, it is interesting to highlight two main points. First, the fine-tuning applied to the pre-trained network requires an amount of manual work proportional to the number of additional annotations used. Second, it is usually not possible to directly optimize the network using the morphological metrics being studied since their calculation is not fully differentiable. Thus, the metric that is optimized is usually the cross-entropy of the class probabilities or the segmentation accuracy, and the morphological metrics are calculated based on the resulting segmentations. As a consequence, there is no guarantee that the original or fine-tuned network will provide segmentations that are unbiased with respect to the desired morphometric analyses. 

To investigate the situation described above, we conducted two main analyses: (i) the identification of possible biases when measuring the morphometry of a blood vessel dataset using an off-the-shelf network, initially optimized to segment a different dataset; (ii) if biases are identified, to which extent they can be mitigated by fine-tuning the off-the-shelf network to the new dataset. We divided our experiments into two steps. First, we optimized two neural networks, each specialized in segmenting either tortuous or non-tortuous blood vessels. Next, we measured the reliability of each specialized network when applied to blood vessels having distinct tortuosity than the blood vessels used during training. 

\section{Material and methods}

\subsection{Measuring the Tortuosity of Blood Vessels}
\label{sec:tortuosity}

Many different metrics for quantifying the tortuosity of blood vessels have been defined in the literature \cite{Smedby1993, Hart1999, Dougherty2000, Grisan2008, Wilson2008}. \textcite{ramos2018retinal} compared the tortuosity scores provided by five experts regarding blood vessels in retina fundus images. They also compared the scores of the experts with those obtained from different automated approaches for measuring tortuosity. They found a high inter-expert variability as well as different degrees of agreement between the experts and the considered automated measurements. Their results demonstrate that there is no optimal approach for defining tortuosity. Indeed, as discussed in \cite{Bullitt2003}, different types of tortuosity can be considered. Therefore, any given tortuosity measurement will have advantages and drawbacks. In this work, we calculate the tortuosity of blood vessels using linear regression residuals~\footnote{Code available at \protect\url{https://github.com/chcomin/pyvane}}. 

\begin{figure}[ht]
    \centering
    \includegraphics[width=0.7\linewidth]{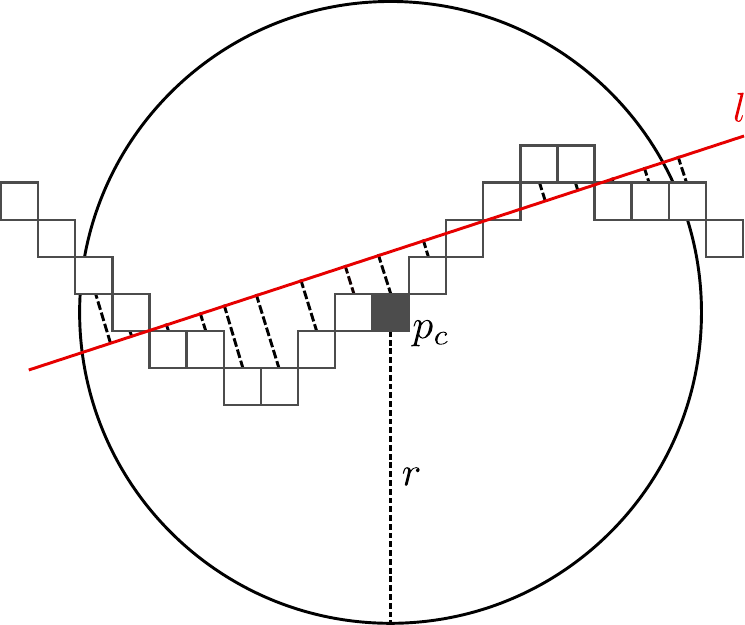}
    \caption{Illustration of the methodology used for calculating the tortuosity of a blood vessel segment. A straight line is adjusted to the points inside a circle of radius $r$ centered at a reference point $p_c$. The tortuosity associated to point $p_c$ is given by the average distance between the line and the points inside the circle.}
    \label{fig:tort_method}
\end{figure}

The tortuosity is obtained as follows. First, the centerlines of the blood vessels are calculated using the Palàgyi-Kuba topological thinning algorithm \cite{Palagyi1998}. Next, we calculate tortuosity maps based on the generated centerlines. As illustrated in Figure \ref{fig:tort_method}, given a reference centerline pixel $p_c$ belonging to a vessel segment, we define a circle of radius $r$ centered at $p_c$. The centerline pixels inside this circle define the neighborhood of $p_c$. Then, a line $l$ is fitted to the set of neighboring pixels using least-squares linear regression. The tortuosity of $p_c$ is then defined as the average of the point-to-line distances between the neighboring pixels of $p_c$ and $l$ (dashed lines in Figure \ref{fig:tort_method}). In other words, the tortuosity is given by the root mean squared error of the least-squares regression of the neighborhood of $p_c$. If the blood vessel segment around $p_c$ is relatively straight, the respective tortuosity will be small. Contrariwise, if the segment cannot be well represented by a straight line, the tortuosity will be large. The overall tortuosity of the blood vessels in an image is then defined as the average tortuosity of the centerline pixels. Notice that the radius $r$ controls the size of the detected tortuous structures. For smaller values of $r$, small sinuous structures will have larger tortuosity values. For larger values of $r$, longer vessels with smoother curvatures will be recognized as more tortuous. Thus, parameter $r$ controls the scale of the analysis. For the experiments in this work, we set $r=10$ since it allows the detection of sharp blood vessel turns in the mouse cortex. An example of tortuosity values obtained for a sample can be found in Figure~\ref{fig:tort_example} of the Supplementary Material.

\subsection{Optimizing a CNN for Blood Vessel Segmentation}
\label{sec:creating_specialized}

The first part of our experiment (illustrated in Figure~\ref{fig:motivation}A) consists in training a CNN capable of segmenting blood vessels with state-of-the-art accuracy. For the sake of generality, we should ideally use methodologies that are well-adopted in the literature. Therefore, our chosen network architecture was the U-Net \cite{Ronnenberger2015}, as it is widely used for segmentation tasks based on small datasets and it performs well on blood vessels \cite{Zhang2018, Jin2019, Livne2019}.


The U-Net is composed of an encoder and a decoder. The encoder maps the input image into feature maps having progressively lower spatial resolutions, while the decoder progressively upsamples the output of the encoder and incorporates feature maps from intermediate stages of the encoder. The idea behind this methodology is to generate pixelwise probabilities for image segmentation while taking into account image features at different spatial resolutions. For our experiments, we use a U-Net with a ResNet-34 encoder \cite{He2016}.

\begin{figure*}
 	\includegraphics[width=\textwidth]{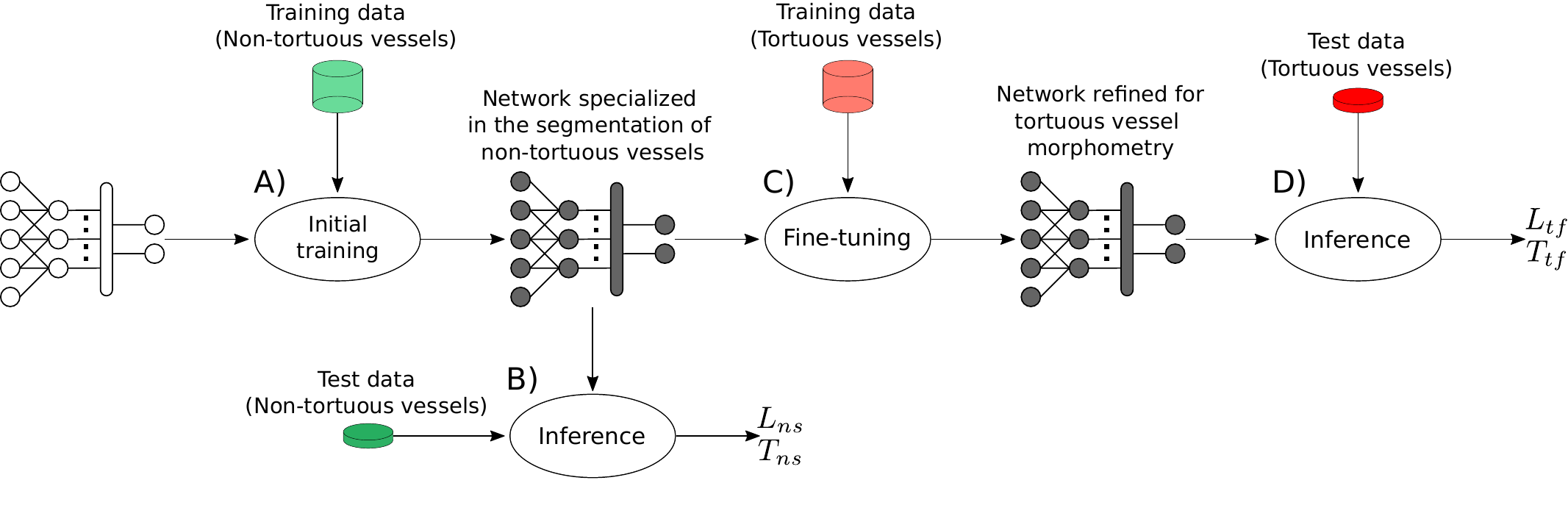}
 	\caption{The methodology used for training and fine-tuning the CNNs considered in the analysis. The case where the network is initially trained on non-tortuous vessels and fine-tuned on tortuous vessels is depicted. The same procedure is used when training on tortuous vessels and fine-tuning on non-tortuous vessels.}
 	\label{fig:refinement_methodology}
\end{figure*}

Recall that the objective of our analysis is to investigate possible biases when segmenting tortuous and non-tortuous blood vessels. Therefore, we optimized two distinct networks. One network is trained on a dataset composed predominantly of tortuous blood vessels, while the other is trained on a dataset where the blood vessels typically have low tortuosity. Section~\ref{sec:dataset} describes the procedure used for generating these two datasets. This initial training step is depicted in Figure~\ref{fig:refinement_methodology}A for the case of non-tortuous blood vessels. This training is performed for 30 epochs. 80\% of the images containing non-tortuous blood vessels were used in the training set, 10\% in the validation set, and 10\% in the test set. The training is carried out using the Adam optimizer \cite{Diederik2017}, with the cross-entropy as a loss function. The learning rate is scheduled using the 1cycle policy \cite{Smith2018}, with a maximum value of 0.0005. The procedure used for training a network on the tortuous blood vessel dataset is identical. Provided the training is successful, the resulting networks will be able to segment blood vessels having similar characteristics as those in the respective training sets. Thus, we henceforth refer to those networks as being \emph{specialized} on tortuous or non-tortuous blood vessels.

As illustrated in Figure \ref{fig:refinement_methodology}B, after training, the network is applied to the test set in order to generate a set of label images $L_{ns}$, where $ns$ stands for \emph{non-tortuous specialized}. From $L_{ns}$, we calculate the average tortuosity value of each image using the methodology described in Section \ref{sec:tortuosity}. The set of calculated tortuosity values is represented as $T_{ns}$. For the network specialized on tortuous blood vessels, we represent the obtained labels as $L_{ts}$ and average tortuosity values as $T_{ts}$, where $ts$ stands for \emph{tortuous specialized}.


\subsection{Fine-Tuning an Off-The-Shelf Network for Blood Vessel Morphometry}
\label{sec:fine-tuning}

The second part of our analysis consists in implementing the process illustrated in Figure~\ref{fig:motivation}B. Thus, the networks trained using the procedure described in the previous section are fine-tuned to a new dataset. The approach used for fine-tuning the networks is illustrated in Figure~\ref{fig:refinement_methodology}C. The figure shows the methodology for the network specialized in non-tortuous blood vessels, but the procedure is identical for the other network. As depicted in the figure, the network is fine-tuned using a subset of the tortuous training set. The fine-tuning is done for 15 epochs. Moreover, we use the same methodology and hyperparameters employed when creating the specialized networks. 

The fine-tuned network is then applied to the test data of the tortuous vessel dataset (Figure~\ref{fig:refinement_methodology}D). A new set of labels $L_{tf}$ and average tortuosity values $T_{tf}$ are generated. $tf$ stands for \emph{tortuous fine-tuned}. As mentioned above, this fine-tuning procedure is also performed for the network specialized in tortuous blood vessels. In this case, we represent the new set of labels as $L_{nf}$ and average tortuosity as $T_{nf}$, where $nf$ stands for \emph{non-tortuous fine-tuned}.


\subsection{Quantifying Morphometry Biases}
\label{sec:morph_quant}

The label images and average tortuosity values obtained from the specialized and fine-tuned networks can be used to assess the segmentation quality and identify possible biases generated by the training procedures regarding the tortuosity of the blood vessels. We define the segmentation quality as the IoU between the labels obtained by the network fine-tuned on a dataset and those obtained by the network specialized on the same dataset. So, for the non-tortuous blood vessels:

\begin{equation}
    IoU_{n} = \frac{1}{N}\sum_{i=1}^{N} IoU(L_{nf}(i), L_{ns}(i)),\label{eq:IoUs}
\end{equation}
where $L_{nf}(i)$ and $L_{ns}(i)$ represent, respectively, the i-th image from the sets $L_{nf}$ and $L_{ns}$, and $N$ is the number of images. The term $IoU(L_{nf}(i), L_{ns}(i))$ in Equation~\ref{eq:IoUs} quantifies the IoU between the labels obtained by a network that was initially trained on tortuous blood vessels and then fine-tuned on non-tortuous blood vessels ($L_{nf}(i)$) and the labels obtained by a network that was specifically trained for segmenting non-tortuous blood vessels ($L_{ns}(i)$). Ideally, $IoU_{n}$ should be close to 1. Notice that this can happen even if the networks cannot segment the blood vessels with good accuracy.  

Equivalently, for tortuous blood vessels:

\begin{equation}
     IoU_{t} = \frac{1}{N}\sum_{i=1}^{N} IoU(L_{tf}(i), L_{ts}(i)).
\end{equation}

The quality of the tortuosity values obtained by each network can also be measured. For the non-tortuous dataset, we define
\begin{equation}
\label{eq:r_tort_s}
    R_{n} = \frac{1}{N}\sum_{i=1}^{N}\frac{T_{nf}(i)}{T_{ns}(i)}.
\end{equation}
The term $T_{nf}(i)/T_{ns}(i)$ quantifies the average tortuosity obtained using the network that was initially trained on tortuous blood vessels and then fine-tuned on non-tortuous vessels ($T_{nf}(i)$) with respect to the tortuosity value obtained using the network specialized on non-tortuous blood vessels ($T_{ns}(i)$).

Similarly, for tortuous blood vessels:
\begin{equation}
\label{eq:r_tort_t}
    R_{t} = \frac{1}{N}\sum_{i=1}^{N}\frac{T_{tf}(i)}{T_{ts}(i)}.
\end{equation}
Values $R_{n}$ and $R_{t}$ quantify the changes in tortuosity obtained when using an off-the-shelf network instead of training a neural network from scratch. It is important to notice that the off-the-shelf network was not trained on completely unrelated data. The tortuous and non-tortuous datasets have similar characteristics, with their main distinction being the tortuosity of the blood vessels contained in the images. Also, given the natural sinuosity of the blood vessels in our images, we observed in our experiments that the values in $T_{ns}$ and $T_{ts}$, being averages calculated over the entire images, did not reach values close to zero. 


Having defined these quality metrics, we search for biases of segmentation and tortuosity values by calculating the values of relative tortuosity and IoU as we increase the number of annotated images used in the fine-tuning step. The number of images was varied from 0 (no fine-tuning) to 40. Since the annotated images are randomly sampled, the fine-tuning is repeated $K$ times, and each quality metric is calculated as the average for all repetitions. Since using less annotated images leads to large fluctuations in the results, we define $K$ as
\begin{equation}
    K = max\left(5, \frac{t}{2n}\right),
    \label{eq:k}
\end{equation}
where $t$ is the total size of the training set and $n$ the number of images used in the refinement. Thus, the smaller the number of images used in the refinement, the larger is the number of repetitions used in the evaluation.

A summary of the metrics defined above is presented in Table \ref{tab:metrics} of the Appendix.

\subsection{Dataset}
\label{sec:dataset}

To apply the methodology proposed in this work, we use a dataset composed of 1838 3D volumes, all obtained from the cerebral cortices of mice. All volumes were collected in 3D using confocal microscopy and have a resolution of $1376 \times 1104 \times 51$ voxels, each voxel representing $0.908\mu m \times 0.908\mu m \times 1\mu m$. The procedure used for obtaining the volumes is described in \citet{Ouellette2020}.

\begin{figure}[ht]
 	\includegraphics[width=\columnwidth]{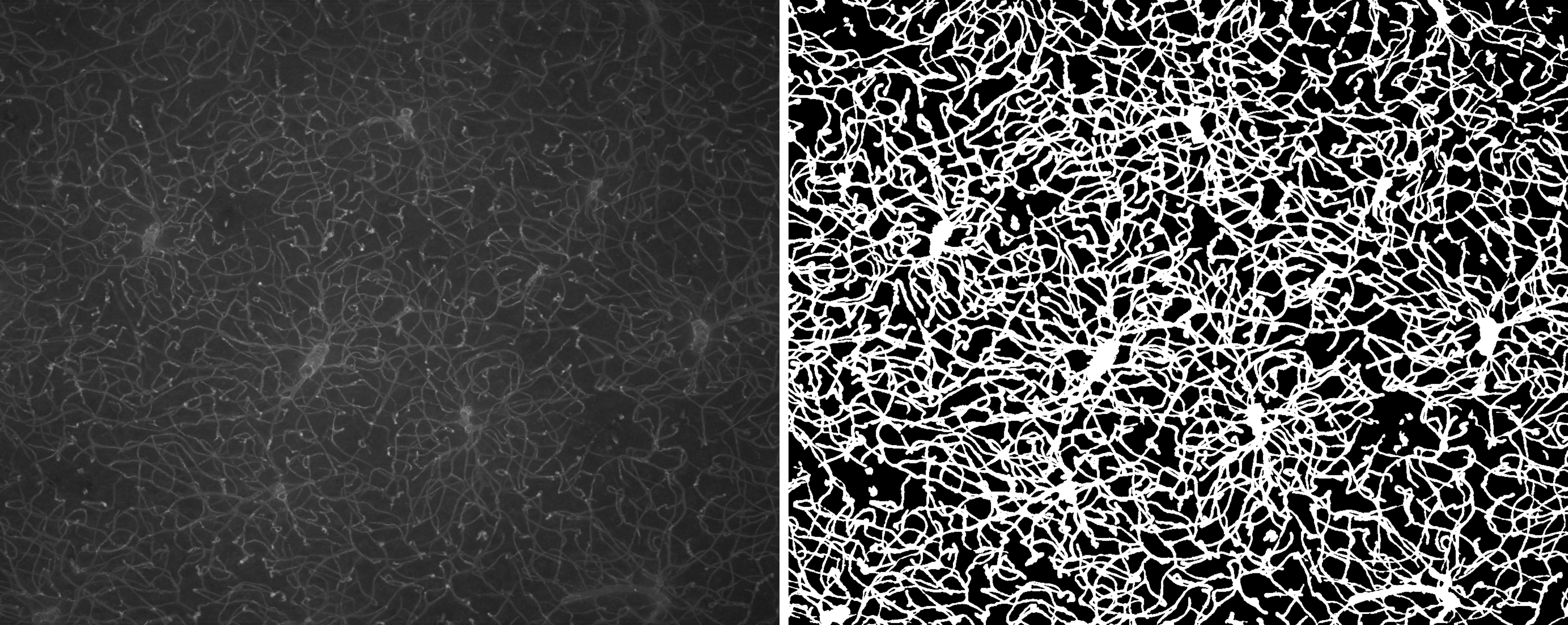}
 	\caption{Labeled cortical vasculature obtained by the semi-supervised segmentation approach.}
 	\label{fig:dataset_example}
\end{figure}

To obtain the ground truth labels, we used a semi-supervised segmentation approach. This same strategy was previously used in other works \cite{Ouellette2020, Lacoste2014}. First, a Gaussian filter with unit standard deviation is applied to each 3D volume. Then, vessel regions are identified by an adaptive thresholding algorithm with a window size of $100 \mu m \times 100\mu m$ applied to each image along the depth of the volume (z-direction). Next, connected components smaller than $500 \mu m^3$ are removed. Since we are interested in evaluating the performance of 2D CNNs, the maximum intensity projections (MIP) of the volumes (along with their respective segmentations) were used as input to the networks -- i.e., 2D images with a resolution of $1376 \times 1104$ pixels. Please refer to \citet{Ouellette2020} for a more detailed description of the methodology. An example of labeled cortical vasculature is depicted in Figure~\ref{fig:dataset_example}.

In order to identify images having mostly tortuous or non-tortuous blood vessels, the average tortuosity value is calculated for each image in the dataset. The 100 images resulting in the highest average tortuosity were labeled as tortuous samples. Similarly, the 100 images with the lowest average tortuosity were labeled as non-tortuous samples. For performance reasons, each image was divided into 16 windows resulting in 1600 samples for each class with a resolution of $344 \times 276$ pixels. We were careful not to include windows from the same image in both training and validation or test sets. All analyses presented in Section~\ref{sec:results} were done using these 3200 samples.


\subsection{Implementation Details}

All experiments were implemented using PyTorch~\footnote{\url{https://pytorch.org/}}. The experiments ran on a desktop computer equipped with an Intel i5-10400f 6 core and 12 threads CPU, 16 GB of RAM, and a Nvidia RTX 2060 6GB GPU. The total processing time of all experiments was approximately 34 days.

\section{Results}
\label{sec:results}

\subsection{Generation of the Specialized Networks}
\label{sec:res_spec_networks}

\begin{figure}[ht]
    \includegraphics[width=\columnwidth]{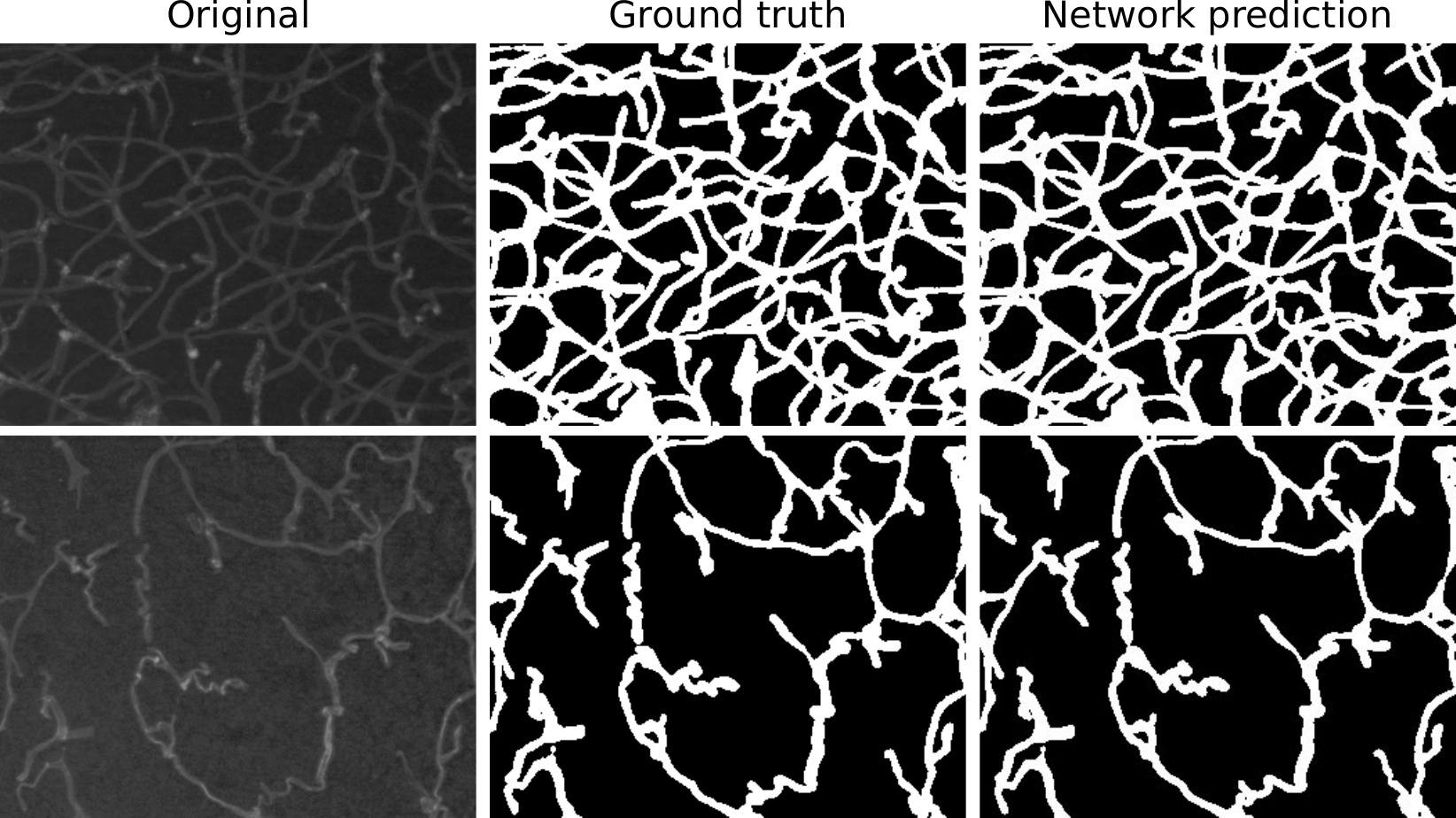}
  \caption{Example of segmentations produced by the specialized networks. Top row: network specialized on non-tortuous vessels applied to an image containing mostly non-tortuous vessels. Bottom row: network specialized on tortuous vessels applied to an image containing mostly tortuous vessels.}
  \label{fig:segmentation_quality}
\end{figure}

Two networks were optimized using the methodology described in Section \ref{sec:creating_specialized}, each specialized in segmenting non-tortuous or tortuous blood vessels. After training, the network specialized in non-tortuous vessels (NSNV) obtained an average IoU of 0.8882 \footnote{Average Dice of 0.9405}. The network specialized in tortuous vessels (NSTV) had an average IoU of 0.9158 \footnote{Average Dice of 0.9553}. These results are comparable to the performance obtained by other works in the literature focused on the segmentation of blood vessels imaged by confocal microscopy \cite{Todorov2020, Tahir2021}. Figure \ref{fig:segmentation_quality} shows examples of results obtained by the CNNs when compared to the ground truth labels.

It is worth mentioning that even though our experiments evaluate the segmentation performance using IoU, we optimized our networks using the cross-entropy as loss function. We observed that using the IoU or Dice as loss functions led to a similar performance compared to the cross-entropy, but the cross-entropy led to better convergence. 

\subsection{Fine-Tuning the Specialized Networks}

\begin{figure}
    \includegraphics[width=\columnwidth]{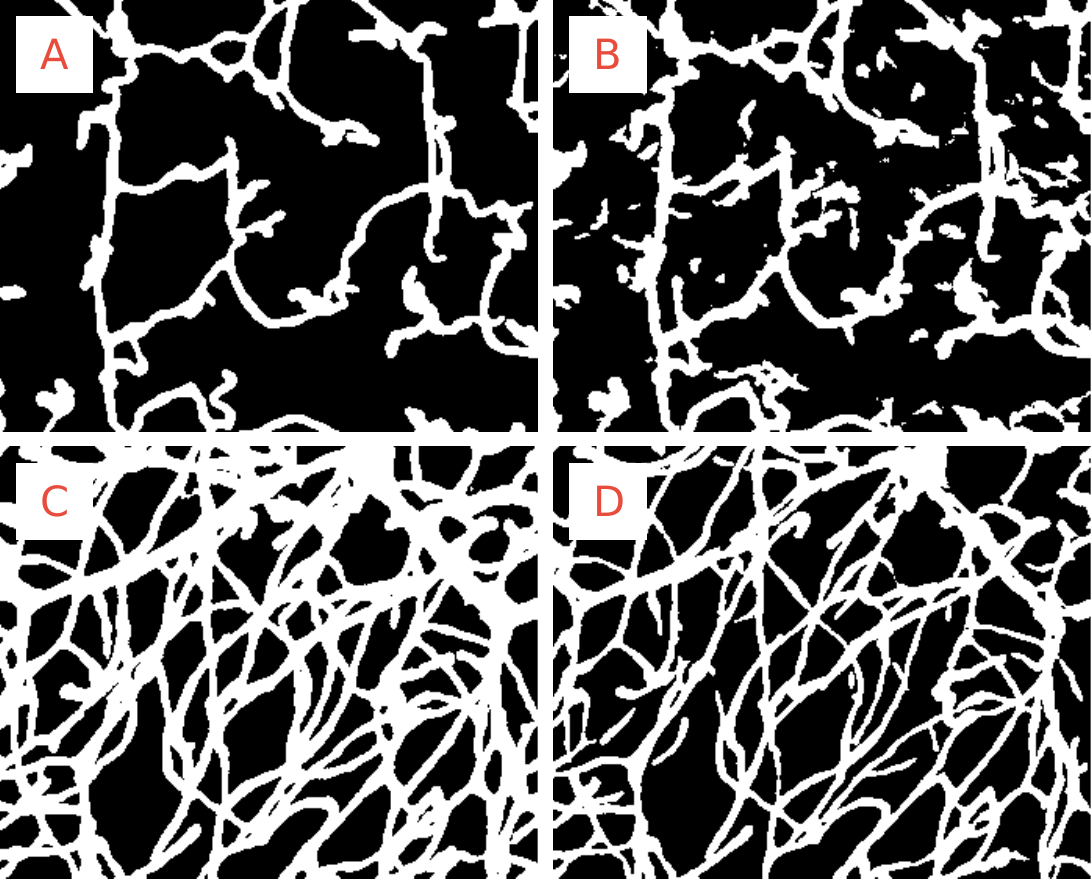}
    \caption{Segmentation artifacts observed when segmenting datasets using inadequate pre-trained networks. (A) Ground truth of a sample containing mostly tortuous blood vessels. (B) Segmentation of the sample shown in (A) using a network trained on non-tortuous blood vessels. (C) Ground truth of a non-tortuous blood vessel image. (D) Segmentation of the sample shown in (C) using a network trained on tortuous blood vessels.}
  \label{fig:biased}
\end{figure}

After optimizing the specialized networks, we examined their ability in segmenting data that they originally were not trained on. When used for segmenting tortuous blood vessels, the NSNV obtained an average IoU of 0.6825. Similarly, when segmenting non-tortuous vessels, the NSTV had an average IoU of 0.6516. Figure \ref{fig:biased} shows examples of typical segmentation results. Figure \ref{fig:biased}B shows the segmentation of blood vessels that are predominantly tortuous obtained using the NSNV. In this case, small connected components that are not observed in the ground truth (Figure \ref{fig:biased}A) are created.  These components represent false positives and thus lead to a high recall and low precision. Such artifacts seem to be associated with the fact that this network tends to classify regions of constant intensity as blood vessels since non-tortuous vessels usually do not present longitudinal intensity discontinuities. Figure \ref{fig:biased}D shows a typical result obtained by the NSTV when applied to images containing blood vessels that are predominantly non-tortuous. By comparing this segmentation with the ground truth (Figure \ref{fig:biased}C), a considerable caliber underestimation is observed, that is, the segmented blood vessels have smaller caliber than the ground truth. Since the result contains many false negatives, the NSTV leads to high precision and low recall when segmenting images with mostly non-tortuous vessels.

The observed segmentation problems can be considered a typical consequence of applying a pre-trained network to a dataset having different statistics than the dataset used for training. Nevertheless, we emphasize that in the present study, the dataset differences are not caused by changes in the sample preparation or imaging protocol but by differences in the morphology of the underlying biological system.

\begin{figure}[ht]
    \centering
    \includegraphics[width=\columnwidth]{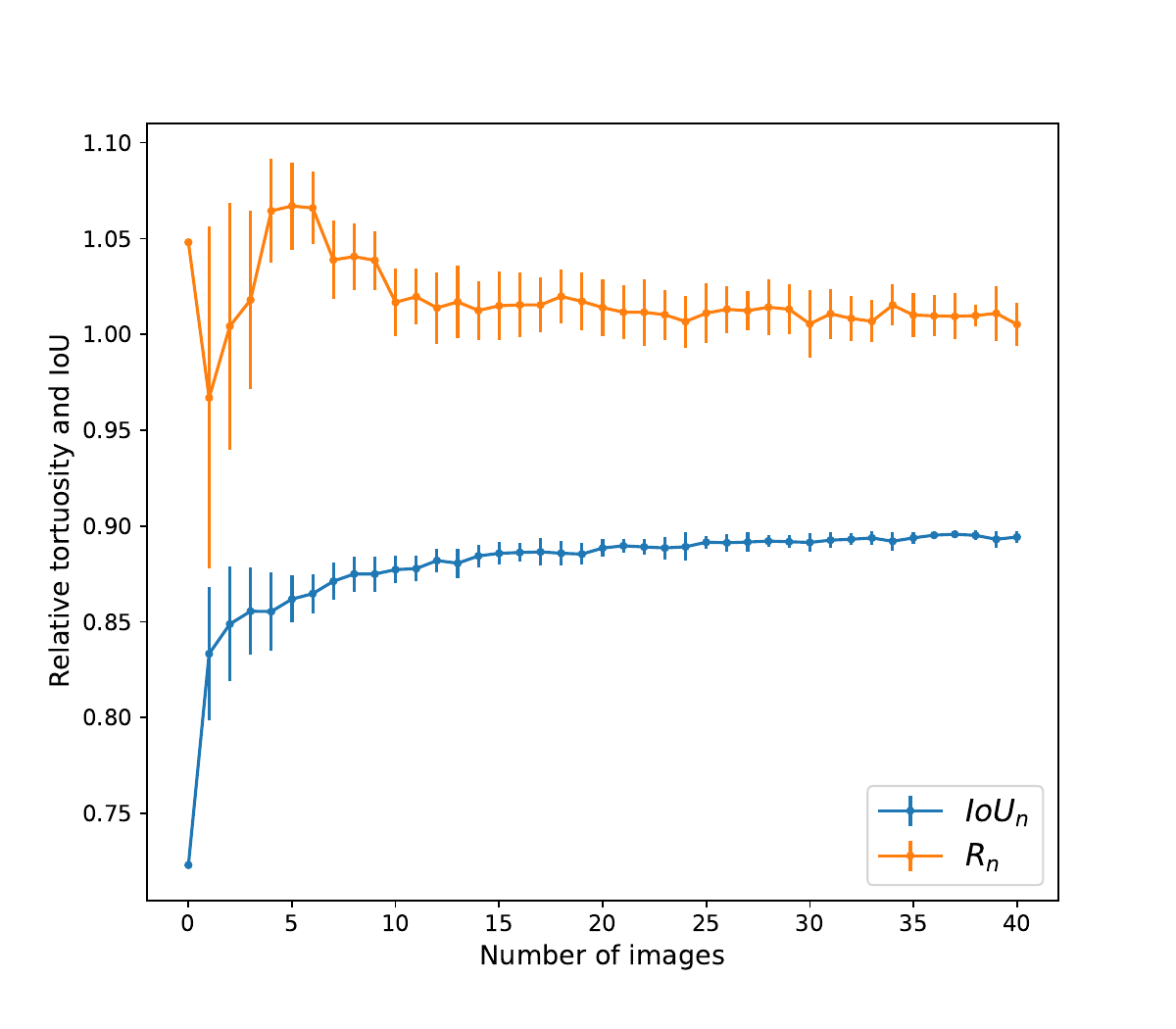}
    \caption{Performance of the network specialized in tortuous vessels when segmenting non-tortuous vessels as a function of the number of non-tortuous images used for fine-tuning the network. The plot shows the average values of $R_n$ and $IoU_n$ obtained after repeating the refining experiment $K$ times (defined in Equation \ref{eq:k}). Error bars depict the standard deviation of $R_n$ and $IoU_n$ over the $K$ repetitions.
    }
    \label{fig:iou_rt_straight}
\end{figure}

Having made the initial characterization of the segmentation problems, we proceed to try to revert the observed biases by fine-tuning the specialized networks. Thus, as presented in Section \ref{sec:fine-tuning}, each specialized network undergoes a refinement step using samples from the dataset where the morphometry analysis is to be performed. Figure \ref{fig:iou_rt_straight} shows the performance of the NSTV when segmenting non-tortuous vessels as a function of the number of non-tortuous images used for fine-tuning. The result shows that the values of $IoU_{n}$ increase with the number of samples. Still, $R_{n}$ does not display the same behavior. The tortuosity is overestimated when the fine-tuning step is performed with up to approximately 13 images. When more images are used, $R_{n}$ stabilizes around 1, while $IoU_{n}$ slowly increases. It is also interesting to note that $IoU_{n}$ and $R_{n}$ are not correlated. Therefore, improvements in a pixelwise accuracy metric, such as the IoU, do not necessarily improve the tortuosity estimation.

\begin{figure}
    \centering
    \includegraphics[width=\columnwidth]{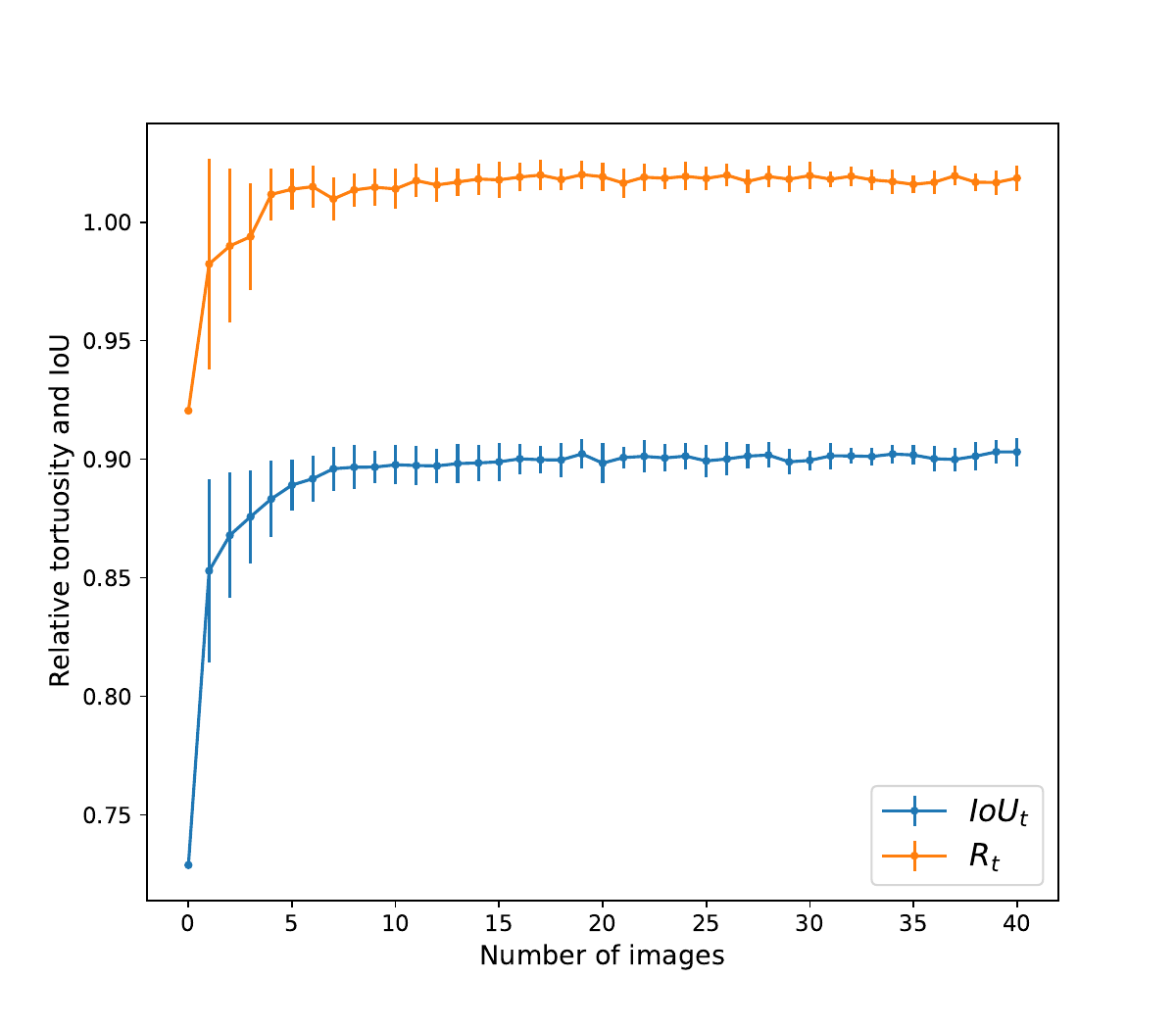}
    \caption{Performance of the network specialized in non-tortuous vessels when segmenting tortuous vessels as a function of the number of tortuous images used for fine-tuning the network. The plot shows the average values of $R_t$ and $IoU_t$ obtained after repeating the refining experiment $K$ times (defined in Equation \ref{eq:k}). Error bars depict the standard deviation of $R_t$ and $IoU_t$ over the $K$ repetitions.}
    \label{fig:iou_rt_tortuous}
\end{figure}

Figure \ref{fig:iou_rt_tortuous} shows the relationship between the performance of the NSNV when segmenting tortuous vessels and the number of images used for fine-tuning the network. As refinement progresses and more tortuous examples are presented to the network, both values of $R_{t}$ and $IoU_{t}$ increase. Also, there is a considerable improvement in the segmentation quality ($IoU_{t}$) when just a few samples are used for fine-tuning. In this case, it was possible to achieve a relative IoU of nearly 0.9 with approximately ten images, in contrast to the 40 images used for achieving the same performance in the previous experiment (Figure \ref{fig:iou_rt_straight}). Non-tortuous vessels usually have simple morphologies. By fine-tuning the network to just a few additional samples containing tortuous vessels, the network quickly adapted to the new morphological characteristics of the samples. Therefore, it was easier to optimize a network for recognizing more complex vessels, starting from a simpler representation, than to optimize it to simplify a more complex vessel representation.

The results obtained from the two experiments considered in this section show that both networks segmented the out-of-distribution samples with similar accuracy. However, each experiment resulted in different degrees of tortuosity underestimation or overestimation depending on the number of images used for fine-tuning. Without fine-tuning, the NSTV overestimated the global tortuosity of the non-tortuous data by approximately $5\%$ compared to the NSNV performance. Similarly, the NSNV underestimated the global tortuosity value of tortuous data by approximately $8\%$ compared to the NSTV performance.  It was observed that these errors could be mitigated by applying a fine-tuning step with 10 to 40 additional images (0.625\% to 2.5\% of the original training set), depending on the tortuosity of the blood vessels.

\section{An approach for improving tortuosity estimation}



The previous experiments have shown that a fine-tuning step involving at least ten images is necessary for partially correcting the morphometry biases of the considered off-the-shelf networks. One drawback of this approach is that manually labeling blood vessels is a burdensome task. Thus, in many situations, annotating a large number of blood vessels may be prohibitive. Furthermore, there is no guarantee that a fine-tuning step with additional samples will work for every dataset and network architecture since it is not clear to which extent our experiments can be generalized. Another difficulty with the fine-tuning approach happens when the off-the-shelf network needs to generalize better towards tortuous vessels. In this regard, tortuous vessels are usually disease markers. Therefore, it becomes harder to optimize a network to increase its generalization capability since more effort is needed to develop, image, and label samples containing tortuous vessels. Thus, it is beneficial to develop approaches for increasing the generalizability of a network using \emph{as few manual labelings as possible}. 



Training CNNs usually requires large amounts of labeled data. This problem motivated the development of several data augmentation techniques over the years. Data augmentation techniques allow a network to learn patterns that are not present in the original data and, consequently, to improve its generalizability. We propose a methodology for increasing the generalization capability of a network specialized in segmenting non-tortuous blood vessels. This is done by applying data augmentation to the set of non-tortuous vessels using elastic transformations \cite{Simard2003}, a type of transformation capable of distorting the blood vessels and giving them a more tortuous appearance.

The elastic transformation deforms an image by drawing random displacement fields. Thus, for each image, horizontal ($\Delta_x$) and vertical ($\Delta_y$) displacement fields are defined by drawing random numbers between -1 and +1, generated from a uniform distribution. Then, the fields $\Delta_x$ and $\Delta_y$ are convolved with a gaussian with standard deviation $\sigma$. In this case, $\sigma$ works as an elasticity coefficient. When $\sigma \approx 0$, the result is an uncorrelated displacement field. For non-zero $\sigma$, an elastic deformation effect is obtained. Lastly, the displacement fields $\Delta_x$ and $\Delta_y$ are multiplied by a scale factor $\alpha$, which controls the transformation intensity. The elastic transformation can generate realistic-looking tortuous vessels based on images containing non-tortuous vessels. An example of the application of elastic transformations in our dataset is depicted in Figure~\ref{fig:ex_elastic_transformation} of the Supplementary Material. Also, Figure~\ref{fig:tort_over_et} of the Supplementary Material shows typical tortuosity values obtained for different values of parameter $\alpha$.

It is important to notice that data augmentation can be applied in two distinct situations. The first involves applying data augmentation when training the network from scratch. The second situation concerns applying data augmentation for fine-tuning a pre-trained network. 

\subsection{Data Augmentation When Training From Scratch}

To evaluate the influence of data augmentation when training a network from scratch, we define a new set of images based on the application of the elastic transformation to the set of non-tortuous vessels, which we henceforth refer to as \emph{false tortuous vessels}. Next, we optimize a neural network using the false tortuous vessels and assess the capability of this network in performing the morphometry of the set of real tortuous vessels. The training methodology employed is the same as the one used in the optimization of the specialized networks (see Section \ref{sec:creating_specialized}), including the architecture, hyperparameters, and the number of epochs. After training, for each segmented image on the test set, the average tortuosity $T_{fs}$ is calculated. $fs$ stands for \emph{false-tortuous specialized}. Similarly to Equations \ref{eq:r_tort_s} and \ref{eq:r_tort_t}, the relative tortuosity in this experiment ($R_{fs}$) is calculated as the average ratio between each element of $T_{fs}$ and the respective element of $T_{ts}$ obtained by the NSTV, that is,

\begin{equation}
    R_{fs} = \frac{1}{N}\sum_{i=1}^{N}\frac{T_{fs}(i)}{T_{ts}(i)}.
\end{equation}
The relative IoU, $IoU_{fs}$, is also calculated. 

\begin{figure}
    \centering
    \includegraphics[width=\columnwidth]{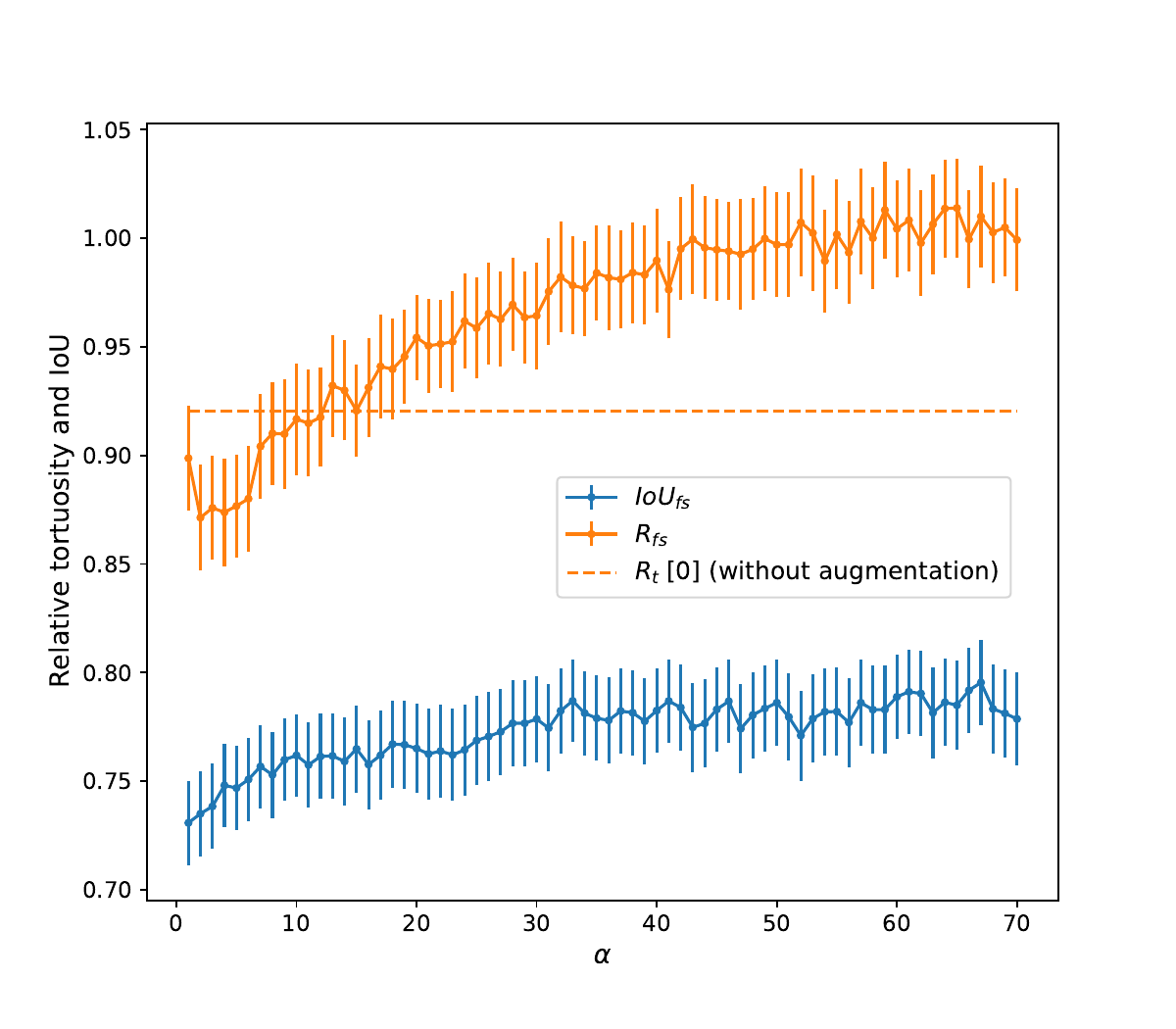}
    \caption{Performance of the network trained with false tortuous vessels when subjected to the morphometry of real tortuous vessels. $\alpha$ is the scale factor used in the elastic transformation to obtain the set of false tortuous vessels. Error bars represent the standard errors of the average values of $R_{fs}$ and $IoU_{fs}$ of the test set. The orange dashed line depicts the base performance of the network specialized in non-tortuous vessels when segmenting tortuous vessels (with no fine-tuning), as can be seen in Figure \ref{fig:iou_rt_tortuous}.}
    \label{fig:rt_da_varying_alpha}
\end{figure}

Figure \ref{fig:rt_da_varying_alpha} displays $R_{fs}$ and $IoU_{fs}$ as a function of the parameter $\alpha$ used in the elastic transformation. The value of $\sigma$ was set to 4. We observed that keeping $\sigma=4$ and varying $\alpha$ between 0 and 70 covered several examples of tortuous blood vessels found in our dataset. There is a considerable drop in $R_{fs}$ when $\alpha$ is between 1 and 10. On the other hand, for $\alpha \geq 12$, training with the set of false tortuous vessels improves the tortuosity estimation. The peak performance is observed when $\alpha$ is around 60, with $R_{fs} \approx 1$. Another important result is the lack of a strong correlation between $R_{fs}$ and $IoU_{fs}$. This reinforces the idea that improvements in vessel morphometry may be caused by effects that are not properly quantified by pixelwise accuracy metrics. 

\subsection{Data Augmentation During Fine-Tuning}

\begin{figure}
    \centering
    \includegraphics[width=\columnwidth]{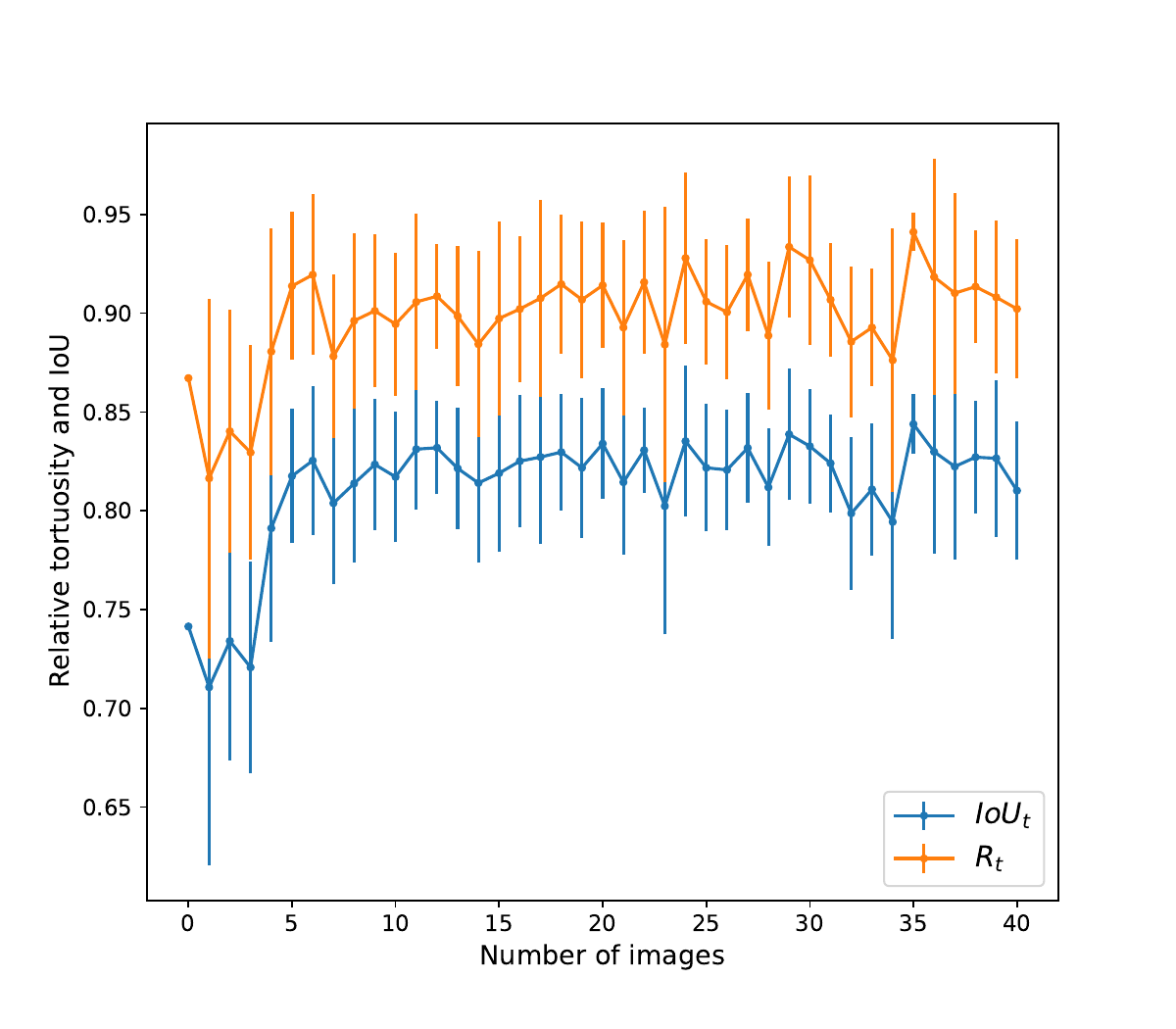}
    \caption{Performance of the network specialized in non-tortuous vessels when segmenting tortuous vessels as it goes through a fine-tuning step with increasingly more false tortuous vessels. The plot shows the average values of $R_t$ and $IoU_t$ obtained after repeating the refining experiment $K$ times (defined in Equation \ref{eq:k}). Error bars depict the standard deviation of $R_t$ and $IoU_t$ over the $K$ repetitions.}
    \label{fig:refinement_behaviour}
\end{figure}

As mentioned above, data augmentation may also be used when fine-tuning a CNN. Thus, it is interesting to verify if the generalizability of a pre-trained network can be increased by fine-tuning the network using false tortuous vessels. This was investigated by fine-tuning the NSNV on false tortuous vessels generated using the elastic transformation with $\alpha=64$ and $\sigma=4$. These parameters were chosen because they provided the best performance in the previous data augmentation experiment. All other parameters were kept the same as those used in the fine-tuning applied to the NSNV (Section \ref{sec:fine-tuning}). The results are depicted in Figure \ref{fig:refinement_behaviour}. It can be observed that there is no noticeable improvement on the values of $IoU_t$ and $R_t$. This result contrasts with the improvement obtained when applying data augmentation for training the network from scratch.

Some considerations can be made from the obtained result. First, data augmentation was unable to correct the bias of the NSNV. In addition, the $R_t$ curve has a large standard deviation, which indicates that the quality of the morphometry obtained by the network is not reliable. Additional approaches could be investigated for fine-tuning the network. Some possibilities include: modifying the learning rate scheduler for finding a solution that better generalizes the data, increasing the number of images used in the fine-tuning step, and using different data augmentation techniques.

\section{Conclusions}

The segmentation and characterization of blood vessels is an important task for diagnosing many types of diseases as well as for studying the development of neurovascular systems. Therefore, defining robust and unbiased approaches for identifying blood vessels in digital images is of great importance. Usually, most of the focus is placed on obtaining good segmentation accuracy, commonly measured using the Dice or IoU metrics. Recent developments of neural networks allowed the definition of highly-accurate segmentation pipelines, which were able to provide results that were close to human-level performance \cite{Todorov2020}. Thus, going forward, we argue that more focus should be given to the influence of the segmentation step on downstream tasks related to blood vessel morphometry. Specifically, when developing computerized systems, it is important to take into account that \emph{a high segmentation performance does not necessarily lead to properly labeled images for morphometric analyses}. This is especially true for neural networks given that explaining the results provided by such systems is still an active area of research \cite{Rudin2019, Adebayo2020}.

Here we focused on identifying to which extent segmentation accuracy is related to morphometric precision when measuring the tortuosity of blood vessels. A procedure was developed to compare a CNN that was trained only on images containing mostly tortuous blood vessels with a CNN that was trained on non-tortuous blood vessels and fine-tuned on tortuous blood vessels. The opposite situation was also investigated. We found that the tortuosity values obtained from a CNN trained from scratch may not agree with those obtained by a CNN that was pre-trained on data acquired from a different experimental condition, and consequently, with slightly different tortuosity statistics. Those biases can be corrected, to some extent, by applying a fine-tuning step using manually labeled data from the new experimental condition. Also, we investigated a possible bias-reducing data augmentation approach using elastic transformations of the training set. This strategy can be helpful for small datasets and also requires fewer manually annotated images. 


One important result of our analysis is that training the network using the proposed data augmentation approach did not lead to a significant improvement of the IoU metric. However, this approach made the network less biased when used for morphometric quantifications. This means that adding proper data augmentation techniques might be useful even if they do not improve segmentation accuracy. Conversely, when using off-the-shelf networks, it might be useful to give preference to networks that were trained with appropriate data augmentation techniques, even if its accuracy is lower than other state-of-the-art architectures. This result also highlights the importance of correctly optimizing a blood vessel morphometry pipeline for prospective studies regarding vascular systems. The optimization should not consider only pointwise accuracy metrics but also the performance of metrics associated with morphological properties of the blood vessels.

For future studies, it would be interesting to consider the influence of segmentation on other morphological quantities, such as blood vessel length and the number of branching points. Also, it is clear that the same analysis considered in this study can be applied to other systems composed of a segmentation step and a downstream quantification task.




\section*{Acknowledgements}
Cesar H. Comin thanks FAPESP (grants no. 18/09125-4 and 21/12354-8) for financial support. The authors acknowledge the support of the Government of Canada's New Frontiers in Research Fund (NFRF) (NFRFE-2019-00641). This study was financed in part by the Coordenação de Aperfeiçoamento de Pessoal de Nível Superior - Brasil (CAPES) - Finance Code 001.

\bibliography{references}

\clearpage
\onecolumngrid
\appendix*
\begin{center}
\section{Performance Metrics Used in This Work}
\end{center}

\renewcommand{\arraystretch}{1.2}

\begin{table*}[!ht]
\caption{\label{tab:metrics} Symbols and corresponding definitions of the metrics used for quantifying the performance of blood vessel segmentation and tortuosity calculation.}
\begin{ruledtabular}
\begin{tabular}{m{0.1\textwidth} p{0.85\textwidth}}
\textbf{Metric} & \textbf{Description}\\ \hline
$L_{ns}$ & Set of labels obtained by the application of the network specialized in non-tortuous vessels on images containing non-tortuous vessels.\\
$T_{ns}$ & Average tortuosity values of $L_{ns}$.\\
$L_{ts}$ & Set of labels obtained by the application of the network specialized in tortuous vessels on images containing tortuous vessels.\\
$T_{ts}$ & Average tortuosity values of $L_{ts}$.\\
$L_{nf}$ & Set of labels obtained by the application of the network specialized in tortuous vessels and fine-tuned on non-tortuous vessels on images containing non-tortuous vessels.\\
$T_{nf}$ & Average tortuosity values of $L_{nf}$.\\
$L_{tf}$ & Set of labels obtained by the application of the network specialized in non-tortuous vessels and fine-tuned on tortuous vessels on images containing tortuous vessels.\\
$T_{tf}$ & Average tortuosity values of $L_{tf}$.\\
$IoU_n$ & Average IoU between the set of labels $L_{nf}$ and $L_{ns}$. \\
$R_n$ & Average ratio between the set of tortuosities $T_{nf}$ and $T_{ns}$. \\
$IoU_t$ & Average IoU between the set of labels $L_{tf}$ and $L_{ts}$. \\
$R_t$ & Average ratio between the set of tortuosities $T_{tf}$ and $T_{ts}$. \\
$L_{fs}$ & Set of labels obtained by the application of the network specialized in false tortuous vessels on images containing tortuous vessels.\\
$T_{fs}$ & Average tortuosity values of $L_{fs}$.\\
$R_{fs}$ & Average ratio between the set of tortuosities $T_{fs}$ and $T_{ts}$. \\
\end{tabular}
\end{ruledtabular}
\end{table*}

\end{document}


\renewcommand\thefigure{S\arabic{figure}} 

\begin{center}
    \textbf{\large Supplementary Material: An Analysis of the Influence of Transfer Learning When Measuring the Tortuosity of Blood Vessels}
\end{center}

\singlespacing

\begin{center}
\section{Vessel Tortuosity}
\end{center}

\begin{figure}[ht]
    \centering
    \includegraphics[width=\linewidth]{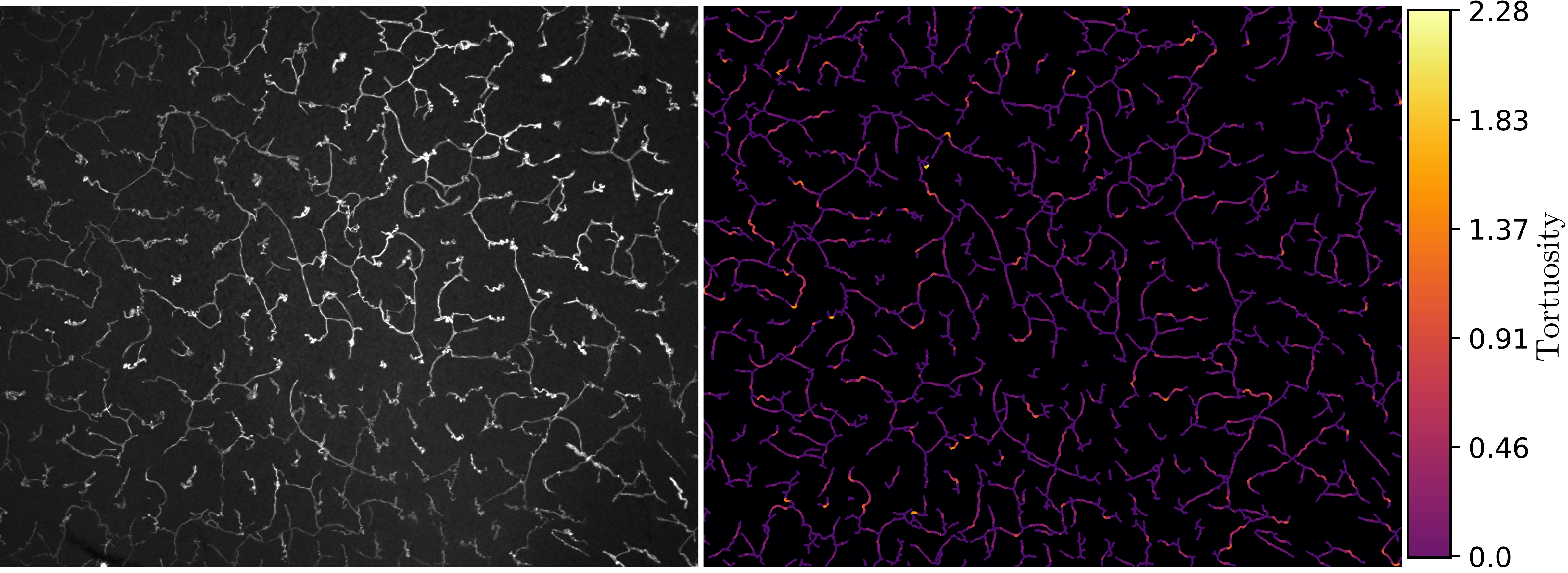}
    \caption{Pixelwise tortuosity obtained for the blood vessels in one of the images of the dataset.}
    \label{fig:tort_example}
\end{figure}

\newpage
\begin{center}
\section{Elastic Transformations}
\end{center}

\begin{figure}[ht]
    \centering
    \includegraphics[width=\linewidth]{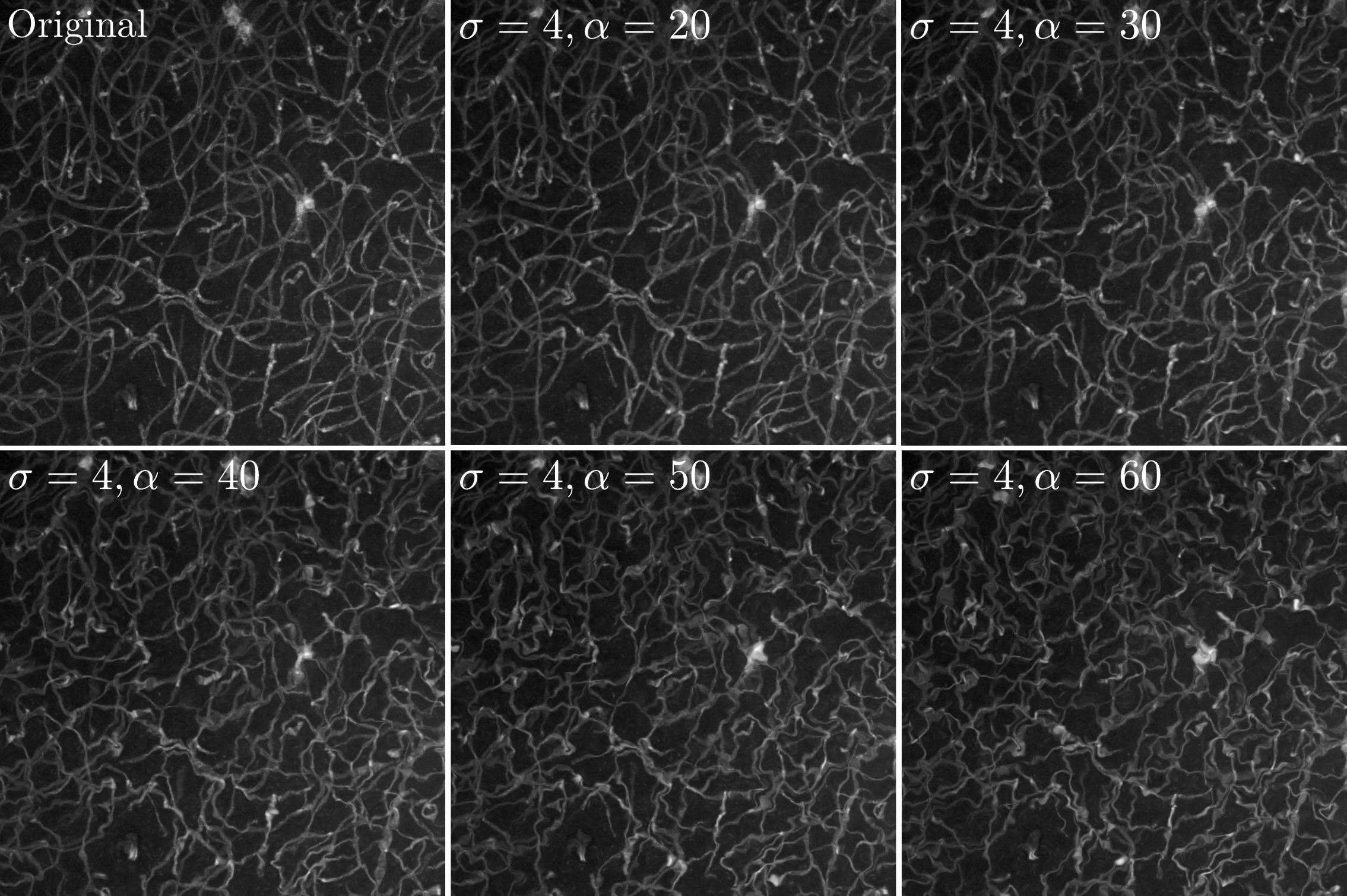}
    \caption{Examples of elastic transformations applied to an image depicting the mouse brain vasculature. The larger the value of $\alpha$, the more tortuous each vessel segment appears to be. 
    }
    \label{fig:ex_elastic_transformation}
\end{figure}

\newpage
\begin{figure}[ht]
    \centering
    \includegraphics[width=\linewidth]{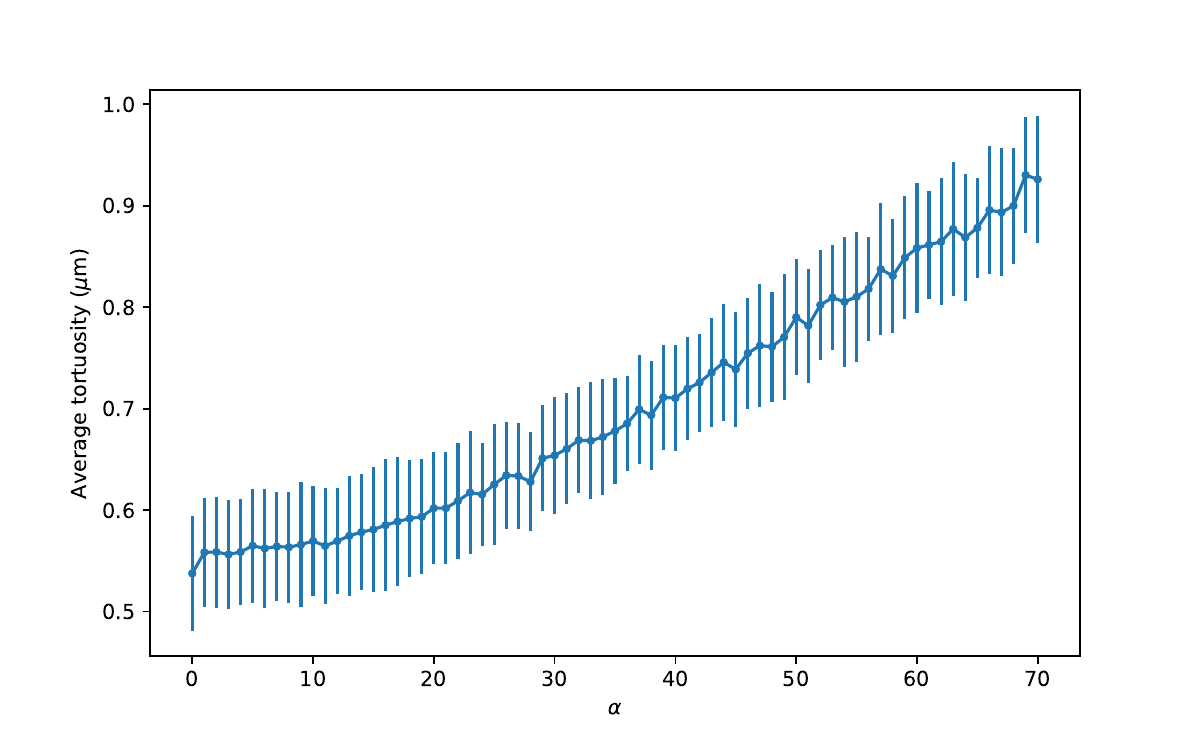}
    \caption{Average tortuosity values of the set of non-tortuous vessels after the application of elastic transformations with different values of $\alpha$. Each point corresponds to the average tortuosity of 50 images randomly sampled from the 1600 images of the dataset containing non-tortuous vessels. The error bars depict the standard deviation of the tortuosity calculated over the 50 images. The same set of images was used for all values of $\alpha$. 
    }
    \label{fig:tort_over_et}
\end{figure}

\newpage